# Exploring Spatial Context: A Comprehensive Bibliography of GWR and MGWR

A. Stewart Fotheringham[1], Chen-Lun Kao[1], Hanchen Yu[2], Sarah Bardin[3], Taylor Oshan[4], Ziqi Li[1], Mehak Sachdeva[1, 5], and Wei Luo[6, 7]

1. Spatial Data Science Center, Department of Geography, College of Social Science and Public Policy, Florida State University, Tallahassee, FL, USA
2. School of Management Science and Real Estate, Chongqing University, Chongqing, China
3. School of Geographical Sciences and Urban Planning, Arizona State University, Tempe, AZ, USA
4. Department of Geographical Sciences, University of Maryland at College Park, College Park, Maryland, USA
5. Department of Urban & Regional Planning, College of Social Science and Public Policy, Florida State University, Tallahassee, FL, USA
6. GeoSpatialX Lab, Department of Geography, National University of Singapore, Singapore
7. Saw Swee Hock School of Public Health, National University of Singapore, Singapore

## 1. Introduction

Whether spatial context affects human behavior has been a topic that has long been debated across the social sciences. Some scholars remain skeptical and argue that spatial context is simply a catch-all term for omitted variables that should be included in models of human behavior. Others contend that there are intangible and unmeasurable effects of places, such as values, beliefs and norms, that need to be accounted for in models of human behavior otherwise the calibration results from such models will be tainted by misspecification biases. Fotheringham and Li (2023), for example, propose that there are two types of contextual effects that should be considered in any model of human behavior. *Intrinsic* contextual effects occur due to the omission of unmeasurable exogenous effects on behavior while *behavioral* contextual effects exist when the conditioned relationship between y and x varies over space. They argue that it is important to include both types of contextual effects in models of human behavior and to assess their importance. Models such as Geographically Weighted Regression (GWR) and Multiscale Geographically Weighted Regression (MGWR) serve as instrumental tools to capture intrinsic contextual effects through the estimates of the local intercepts and behavioral contextual effects through estimates of the local slope parameters (Fotheringham *et al.* 2024).

## 2.1 Geographically Weighted Regression

Geographically Weighted Regression (GWR) was first proposed by Fotheringham *et al* (1996) and Brunsdon *et al.* (1996) and popularized in Fotheringham *et al*. (2002). Refining the global Ordinary Least Squares (OLS) regression framework to allow for spatially heterogeneous processes, GWR is specified as follows:

$$y_i = \beta_{0i} + \beta_{1i}x_{1i} + \beta_{2i}x_{2i} + \ldots + \beta_{ki}x_{ki} + \varepsilon_i \quad (1)$$

where $y_i$ is the value of the dependent variable at location $i$, $\beta_{0i}$ is the local intercept, $x_{ki}$ is the observation of the $k^{\text{th}}$ explanatory variable at location $i$, $\beta_{ki}$ is the $k^{\text{th}}$ coefficient estimate calibrated for location *i,* and $\varepsilon_i$ is the random error term for $i$ = {1, 2, 3,…, n}. GWR borrows data from nearby locations through spatial weights $\boldsymbol{W}$ specified by a *priori* kernel function with a bandwidth optimized from the data to perform a local regression at each location. One major shortcoming of the GWR model, however, is that it assumes that each relationship modeled varies across space at the same scale; a limitation that is overcome by Multiscale Geographically Weighted Regression (MGWR) (Fotheringham et al., 2017; 2024).

## 2.2 Multiscale Geographically Weighted Regression

Unlike GWR, MGWR allows different processes to vary over space at different spatial scales so that some processes could be modeled as being constant over space while others could vary rapidly or slowly over space. MGWR is specified as follows:

$$y_i = bw_0(\beta_{0i}) + bw_1(\beta_{1i}x_{1i}) + bw_2(\beta_{2i}x_{2i}) + \ldots + bw_k(\beta_{ki}x_{ki}) + \varepsilon_i \quad (2)$$

where $bw_0$, $bw_1$, $bw_2$,..., $bw_k$ are the variable-specific bandwidths used to parametrize the data-borrowing schemes to model spatial varying processes, $y_i$ is the dependent variable at the location $i$, $\beta_{0i}$ is the local intercept, $x_{ki}$ is the observation of the $k^{th}$ explanatory variable at the location $i^{th}$, $\beta_{ki}$ is the $k^{th}$ coefficient estimate, and $\varepsilon_i$ is the random error term for $i = \{1, 2, 3, \ldots, n\}$. By allowing covariate-specific bandwidths to be optimized in the model calibration, MGWR can capture globally, and locally varying spatial processes simultaneously and further separate intrinsic contextual effects via the estimates of the local intercept.

## 3. Bibliography

While GWR and MGWR have primarily been utilized to explore spatial contexts within social sciences, the empirical applications extend well beyond this field. There have been thousands of applications and model enhancements across a wide range of disciplines since the seminal proposal of GWR. MGWR gained further recognition by enhancing the scalability and accuracy of the local regression-based framework for spatial data analysis. The use of both GWR and MGWR have been aided by freely available, user-friendly, software available from the following site: https://cosspp.fsu.edu/sdsc/mgwr/ (Oshan *et al.* 2019).

This nearly 500-page bibliography aims to serve as a comprehensive compilation of peer-reviewed papers that have utilized GWR or MGWR as a primary analytical method to conduct spatial analyses in various disciplines. These studies are categorized into fourteen major fields in the bibliography: Agriculture, Archaeology, Cartography and Geovisualization, Community study, Crime, Digital Elevation Models (DEM), Demographics, Dialects, Economics, Ecosystem, Education, Energy, Environmental study, Fire, Fisheries, Floods, Forestry, Geology, Health, Land Use, Landslides, Methodology, Politics, Real Estate, Regional analysis, Software development,

Terrorism, Transportation, Urban study, and Vegetation. Researchers in agricultural science, for example, have been using these methods to analyze crop yield and agricultural policy. GWR and MGWR have become prevalent not only in human geography but in natural geographical sciences, analyzing phenomena such as PM2.5, landside detection, water conservation, urban heat and climate change. Since the outbreak of COVID-19, studies in public health have been using GWR and MGWR to capture heterogeneous spatial processes. Political scientists have used the models to uncover locational influences on voting behaviors. Similarly, the disparities in housing prices have also been explained by spatially varying spatial processes. The widespread adoptions of these papers have shown that GWR and MGWR provide simple implementation yet powerful frameworks that could be extended to various disciplines that handle spatial data. This bibliography therefore serves as a repository and reference site for papers relating to local regression-based modeling and acts as a useful guide to anyone searching the literature for previous examples of local statistical modeling in a wide variety of application fields.

## Agriculture:

## Archaeology:

## Cartography and Geovisualization:

## Community:

## Crime:

## DEM:

## Demographics:

## Dialect:

## Economics:

## Ecosystem:

## Education:

## Energy:

## Environment:

## Fire:

## Fisheries:

## Flood:

## Forestry:

## Geology:

## Health:

## Land Use:

## Landslide:

## Methodology:

## Politics:

## Real Estate:

## Regional Analysis:

## Software:

## Terrorism:

## Transportation:

## Urban Studies:

## Vegetation: